# THE IMPACT OF PHARMACYBERNETIC IN REDUCING MEDICATION ERROR

Dr. Muhammad Shahzad Aslam

Department of Pharmacy, Bahauddin Zakariya University, Multan, Pakistan
Muhammad.shahzad.aslam@hotmail.com

## ABSTRACT

*Doctors and Pharmacists play a foremost role in safe, effective use of medication in health care. Still, there is no database available through which Doctor can communicate with all field of pharmacy such as hospital Pharmacy, Clinical Pharmacy, Community Pharmacy, Nutrition Pharmacy and Drug research center so that they would like to cooperate with pharmacists in Medication error prevention, Drug-Disease management, Nutrition management, and pharmacotherapy. The authors examined the comprehensive project of implementing Electronic Drug Information Record (EDIR), introduce the new term Pharmacybernetic and how to reduce the medication error by integrated management system (IMS). This paper presented EDIR conceptual model and the flow sheet of the Pharmacybernetic system, which describes the integration of different Pharmaceutical related aspect in the field of Cybernetic.*

## 1. INTRODUCTION

Pharmacybernetic is the branch of cybernetic that deals with the study of control and communication of drug. Electronic Drug information record (EDIR) will create database that consist of all the information about Drug. This will help for the scientist for new drug development through which we can easily search of new drugs either from natural, semisynthetic or from synthetic source. It is also beneficial for drug modification having drug solubility, poor bioavailability and other physicochemical parameter. We can also integrate the information about different drugs interaction, drug food interaction, and toxicological data of different drugs, therapeutic drugs comparison and new drug combinations of already resistant diseases. Pharmacist can up-to-date his knowledge through electronic drug information record (EDIR). Physicians and Pharmacists play a major role in safe use of medication in primary outpatient care. Still, little is known about primary care physicians' perceptions of medication errors and error prevention and how they would like to cooperate with pharmacists in error prevention and management [1]

### 1.1 Mission of Pharmacybernetic:

Pharmacybernetic is supposed to enable essential development to the EDIR. EDIR will considerably improve the process information containing collect, organize, extend, and utilize. Until now, the several of healthcare records were printed on paper [2]. The EDIR will help the data to reorganize and update and we can easily access by our computers or on mobiles. The Pharmacybernetic open up inventive possibility while it allows various people to view the same record at the same time from different computers, also to improve the most updated information.

## 1.2 Advantages of Pharmacybernetic:

EDIR has a great acceptability and it is a secure procedure for saving information. It helps to see all pharmacological drug action and it source in one platform. It helps the Pharmacist to search new sources for treatment of diseases and also modify the already available drugs in the market. For the last thirty years, the enlargement and operation of computerized has been observing as a really hard task. Now, we are introducing EDIR which is supposed to permit enlarged uniformity and decreased redundancy of information [3]. The better use of this technology has allowed pharmacist, doctors, nurses, and clinical administrators to perform tasks faster and easier [4].

## 1.3 Problem Statement

There are no integrated foundations of Drug information which contain not only the clinical parameters but also drug design of single drug. Number of new drugs is approved consistently by FDA but their record is not maintained electronically. There is no electronic record of poison of natural source or by chemical source. Doctors are not up-to-date with new medicine since they are using traditional medicine. Use of different drug guide by the pharmacist or doctor in emergency case is time taken and patient need immediate treatment. There are no records of the drugs that are obtained from several sources. It is not possible to remember the dose and drug interaction of all drugs. There is no communication between pharmacist and doctor through which doctor can express about the after-math of drugs and through which pharmacist can work on the drug to reduce its toxicity or to decrease it Adverse side effect. There is the big gap between researcher and Doctor. As a result patient will suffer. There are no mechanisms through which Pharmacist, Doctors and researcher will work on single platform to solve problem related drugs.

## 1.4 Research Objective

The purpose of this research is to propose a new model for Electronic Drug information record (EDIR) application. It will recommend a model to improve the Drug Information System (DIC). The Objective of this research is to ensure sufficient in-depth knowledge of the challenges that we are facing related to Drug database, particularly in the Drug research center (DRC) and how to develop EDIR infrastructure and the use of cybernetic in the field of Pharmacy. In this research the specific objectives are as follows:

• To study detail understanding of electronic drug information record, different medication & Pharmaceutical Error and does detail literature review.

• To propose a new electronic drug information record model, improve the Drug Information System (DIC) and Drug research center (DRC)

• To integrate cybernetic in the field of Pharmacy.

## 2. LITERATURE REVIEW

Information and Communication Technology (ICT) has become the information resource of both range and prerequisite and has thus inspired from the boundary of healthcare. Swift developments in ICT with reduced costs, upgraded trustworthiness and better robustness are enabling a new wave of transform in how and where healthcare can be delivered. Many researchers consider that the electronic record will noticeably modify healthcare, rather than simply switching the Paper-based Record (PR). This modification permits data to be used for a extensive diversity of purposes ranging from direct patient care, decision maintain, quality promise, scientific research, and management of healthcare facilities [5]. The bio cybernetic loop [6] is the essential component of a physiological computing system. The loop functions as a conceptual entity derived from control theory [7] that also defines the flow of data within the system. From bio cybernetics, we amend this phenomenon in the arena of pharmacy and gives the term Pharmacybernetic. Pharmacybernetic is an upcoming zone of pharmacy which comprises highly advanced skills and expertise to deal with technologies in relation to medicines and drugs. The term 'pharma' is derived from the Greek term 'pharmakon' meaning drugs or poisons [8], and 'cybernetics' comes from the Greek term 'kubernetes', which can be translated to mean 'the art of steering' [9]. The concept of 'pharma-cybernetics' is familiarized through the formation of an cooperative tool which consists of a pill-catching game and hangman game planned to assist users to learn about warfarin tablet strengths and drug interactions, based on user-centered (UCD), experience-centered (ECD), and activity-centered design (ACD) approaches [10]. EDIR is an enormous record about drug chemistry, pharmacokinetic, pharmacodynamic, toxicological and clinical data and there are many efforts in the past to merge into single electronic database center which is beneficial not only for the pharmacist, doctor or researcher but also help to the humanity. A methodological review of a user-centered structure that integrates best practices in literacy, information quality, and human–computer interface design and evaluation to guide the design and redesign process of a consumer health website. Following the description of the methods, a case analysis is presented, demonstrating the successful application of the model in the redesign of a consumer health information website with call center. Assessments between the iterative revisions of the website showed improvements in usability, readability, and user satisfaction [11]. Stephanie Holmgren [12] had work in the Toxicology Data and Information Management .Yu HsuanYen [13] studied the comparison of the efficiency and influence of an electronic ADE management system with a traditional working model at a medical center. Electronic health records (EHRs) have increased in popularity in many countries. Pushed by legal mandates, EHR systems have seen substantial progress recently, including increasing acceptance of standards, enhanced medical terminologies and improvements in technical infrastructure for data sharing across healthcare providers. Although the progress is directly helpful to patient care in a hospital or clinical setting, it can also support drug discovery. EHR is helpful in a drug discovery framework: finding novel relationships between diseases, re-evaluating drug usage and discovering phenotype–genotype associations. We believe that in the near future EHR systems and associated databases will effect considerably how we discover and develop safe and efficacious medicines [14]. The continually rising number of chemical compounds and uninterruptedly growing amount of data associated with them render it impossible to handle this information using the manual, traditional techniques that have been used for the last two hundred years of chemical sciences. With the birth of the computer it was undoubtedly accepted that the diversified data in chemistry could only be handled by electronic means. An impressive number of electronic sources are now available, and there is an almost bewildering number of versions and platforms (hosts, in-house systems, etc.) for major sources. Add to this the continuing changes in all the data collected, and this chapter can only provide an orientation and a starting point for further enquiries. T. Engel gives a flavor of the most essential databases used for drug discovery. Also, he focuses on chemical databases; sequence and other biological and biochemical databases are discussed in detail [15]. Clinical event monitors are a kind of lively medication monitoring system that can use signals to alert clinicians to possible adverse drug reactions. The primary goal was to estimate the positive predictive values of select signals used to systematize the detection of ADRs in the medical intensive care unit. [16]

## 3. RESEARCH METHODOLOGY

Electronic Drug Information Record (EDIR) diverges largely with lots of groups centering on the provider requirements for information management and information sharing. The basic purpose to integrate the pharmaceutical services is to provide better health care to the patient and to build stronger relationship between doctor and pharmacist.

### 3.1. Purposed Model of Electronic Drug Information Record

The technological adaptation of Pharmaceutical records is the need of the hour. So we develop a database that integrates the information of all kind of drug. As below in Figure 1 we have shown proposed Electronic Drug Information Record (EDIR).

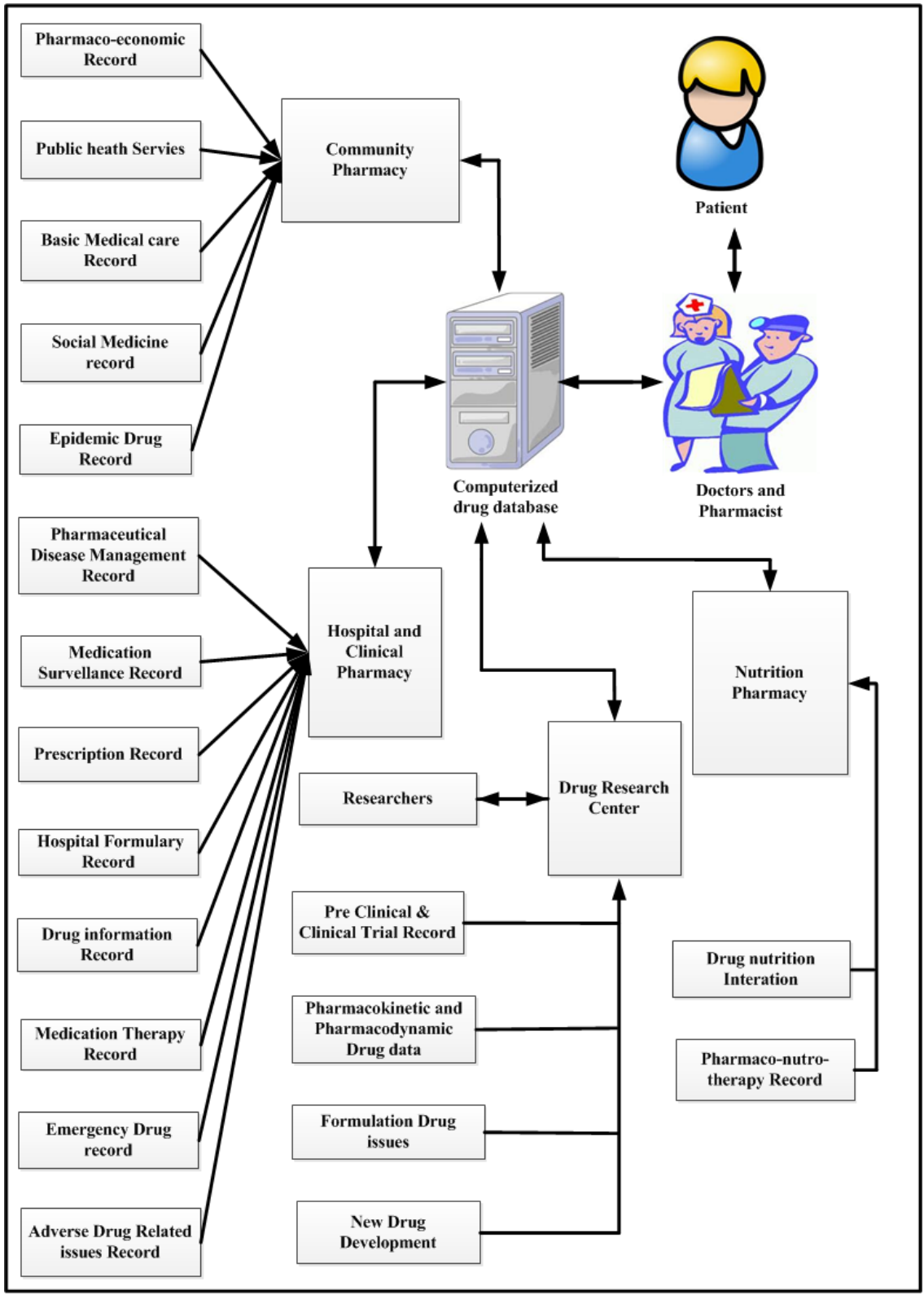

Figure 1. Electronic Drug Information Record

### 3.2. Flow sheet of cybernetic system

Cybernetic system consist of integrated database providing data from different Health and pharmaceutical related field such as Public health record and Epidemic Drug record from the community pharmacy, Drug event monitoring in hospitals, Prescription and pharmaconutrotherapy record from the hospitals and clinics. We can collect the data from the patient in hospitals, clinics and community pharmacy. The record then reach to drug research center where it is collected in the drug database which can easily access to the doctors and pharmacist.

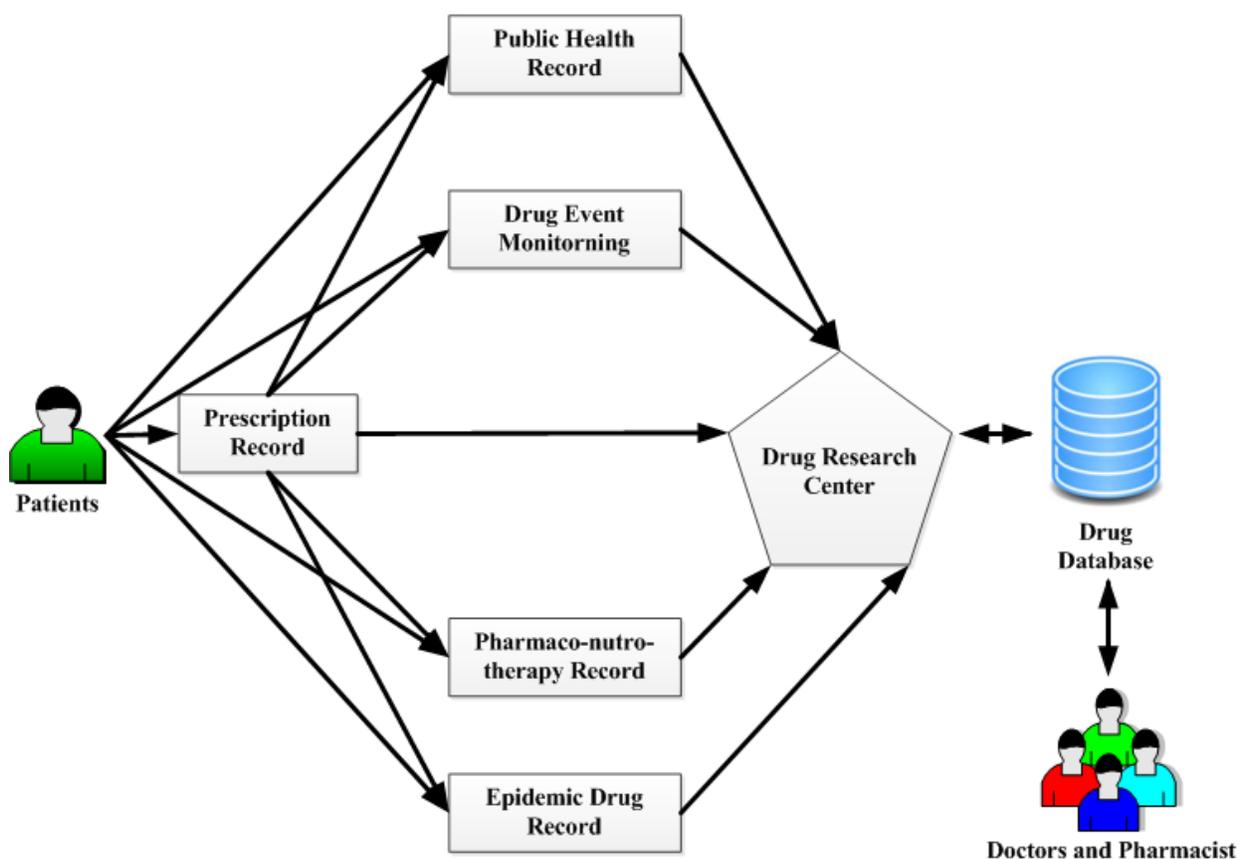

Figure 2.   Flow sheet of Pharmacybernetic System

All the record will enter into drug reach center and drug database. Drug research center will receive all the enquiry related to prescription, Drug event monitoring, nutritional consultancy and Epidemic Drug record and researcher work on new drug development, Drug Pharmacokinetic and Drug Pharmacodynamic parameter. This will help Pharmacist and doctor to identify medication error.

## 4. CONCLUSIONS AND FINDING

The outcome of this research is to recommend a new model that supports the Electronic Drug information record (EDIR) application. It will study indispensable related work to study the accessible Public health record, Drug event monitoring, Prescription Monitoring, Pharmaconutrotherapy and Epidemic Drug record. It will propose new model to improve the Drug Research Center. The objective of this research is to ensure satisfactory EDIR framework and integrate Cybernetic in the field of Pharmacy to meet future challenges

## ACKNOWLEDGEMENTS

I would like to thanks to my Parents. Without their support it could never be possible.

## Authors

Dr. Muhammad Shahzad Aslam obtained the Doctor of Pharmacy (PHARM-D) Degree from the Baqai Medical University, Karachi, Pakistan. He is currently doing Master of Philosophy (M.Phil.) from Bahauddin Zakariya University,Multan, Pakistan in Pharmaceutical Chemistry

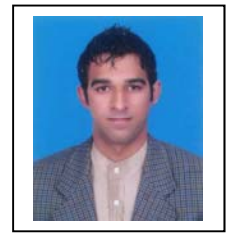